\documentclass[a4paper,12pt]{article}
\usepackage[cp1251]{inputenc}
\usepackage{graphics}
\usepackage{amssymb,amsmath,latexsym}
\usepackage[english]{babel}
\usepackage{indentfirst}
\usepackage{graphicx,times}
\usepackage{natbib}
\begin{document}

\bigskip
\centerline {\bf Correlation Study Of Some Solar Activity Indices}
\centerline {\bf In The Cycles 21 - 23}
\bigskip

\centerline {E.A. Bruevich $^{a}$ , V.V. Bruevich $^{b}$, G.V.
Yakunina $^{c}$ }

\centerline {\it $^{a,b,c}$ Sternberg Astronomical Institute, Moscow
State
 University,}
\centerline {\it Universitetsky pr., 13, Moscow 119992, Russia}\

\centerline {\it e-mail:  $^a${red-field@yandex.ru},
$^b${brouev@sai.msu.ru}, $^c${yakunina@sai.msu.ru}}\

\bigskip
{\bf Abstract.} The correlation coefficients of the linear
regression of six solar indices versus $10.7$ cm radio flux
$F_{10.7}$ were analyzed in solar cycles 21, 22 and 23. We also
analyzed the interconnection between these indices and $F_{10.7}$
with help of the approximation by the polynomials of second order.
The indices we've studied in this paper are: relative sunspot
numbers - SSN, $530.3$ nm coronal line flux - $F_{530}$, the total
solar irradiance - TSI, Mg II $280$ nm core-to-wing ratio UV-index,
Flare Index - FI and Counts of flares. In the most cases the
regressions of these solar indices versus $F_{10.7}$ are close to
the linear regression except the moments of time near to the
minimums and maximums of 11-year activity.
 For the linear regressions we found that the minimum values of
 correlation coefficients $K_{corr}(t)$ for the solar
indices versus $F_{10.7}$ and SSN occurred twice during the 11-year
cycle.

\bigskip
{\it Key words.}  Solar cycle: observations, solar activity indices.

\vskip12pt
\centerline
{\bf1. Introduction}
\vskip12pt

Magnetic activity of the Sun is called the complex of
electromagnetic and hydrodynamic processes in the solar atmosphere.
The analysis of active regions (plages and spots in the photosphere,
flocculae in the chromosphere and prominences in the corona of the
Sun) is required to study the magnetic field of the Sun and the
physics of magnetic activity. This task is of fundamental importance
for astrophysics of the Sun and the stars. Its applied meaning is
connected with the influence of solar active processes on the
Earth's magnetic field.

We have studied monthly averaged values of six global solar activity
indices in magnetic activity cycles 21, 22 and 23. Most of these
observed data we used in our paper were published in
(Solar-Geophysical Data Reports 2009) and (National Geophysical Data
Center. Solar Data Service 2013).

 All the indices
studied in this paper are very important not only for analysis of
solar radiation formed on the different altitudes of solar
atmosphere, but for solar-terrestrial relationships as the key
factors of the solar radiation influence (Extreme Ultra Violet and
Ultra Violet EUV/UV)solar radiation is the most important) on the
different layers of terrestrial atmosphere also.

It's known that the various parameters of solar activity correlate
quite well with the popular sunspot number index - SSN (or relative
sunspot numbers) and with each others over long time scales. (Floyd
{\it et al.} 2005) showed that the mutual relation between sunspot
numbers and three solar UV/EUV indices, the $F_{10.7}$ flux and the
Mg II core-to-wing ratio, which is the important chromospheric UV
index (Viereck {\it et al.} 2001), Viereck {\it et al.} 2004),
remained stable for 25 years until 2000. At the end of 2001 these
mutual relations dramatically changed due to a large enhancement
which took place after actual sunspot maximum of the cycle 23 and
the subsequent relative quietness intermediate called the Gnevyshev
gap.

In our issue for all the indices we used the monthly averages
values. Such averages allowed us to take into consideration the fact
that the major modulation of solar indexes contains a periodicity of
about 27-28 days (corresponding to the mean solar rotation period).
So we reduced the influence of the rotational modulation of the data
sets.

(Vitinsky {\it et al.} 1986) have analyzed solar cycles 18 - 20 and
pointed out that correlation for relative sunspot numbers versus
radio flux $F_{10.7}$ does not show the close linear connection
during all the activity cycle. Also it was emphasized the importance
of statistical study in our solar activity processes understanding.
To achieve the best agreement in approximation of spot numbers
values by $F_{10.7}$ observations (Vitinsky {\it et al.} 1986)
proposed to approximate the dependence SSN - $F_{10.7}$ by two
linear regressions: the first one - for the low solar activity
(where $F_{10.7}$ less than 150 $sfu$) and the second one - for the
high activity ($F_{10.7}$ more than 150 $sfu$). A solar flux unit
$(sfu) = 10^{-22} \cdot W \cdot m^{-2} \cdot Hz^{-1}$

In this paper we found out that the linear correlation was violated
not only for maximums of solar activity cycles but for minimums of
the cycles too. Our analysis of the interconnection between these
indices and $F_{10.7}$ with help of the approximation by the
polynomials of second order confirmed this fact.

We also analyzed the dependence of time of a three-year determined
correlation coefficients $K_{corr}(t)$ (in linear regression
assumption) for solar activity indices versus $F_{10.7}$ and versus
SSN.

Since the nature of solar activity is very complex so we have to
keep in mind that there are different sources which give the
contribution to the value of the indices studied in this paper.

The magnetic activity of the Sun is called the complex of
electromagnetic and hydrodynamic processes in the solar atmosphere
and in the underphotospheric convective zone, see Rozgacheva \&
Bruevich 2002. The analysis of active regions (plages and spots in
the photosphere, flocculae in the chromosphere and prominences in
the corona) requires to study the magnetic field of the Sun and the
physics of magnetic activity. This task is of fundamental importance
for astrophysics of the Sun and the stars. Its applied meaning is
connected with the influence of solar active processes on the
Earth's magnetic field.

For solar energy  coming to  the atmosphere of the Earth, it is
desirable to have solar indices and proxies that vary differently
through time. This strategy of using multiple solar indices has
significantly improved the accuracy of density modeling of the
atmosphere of the Earth has reported by (Bowman, {\it et al.} 2008).
Use of these solar indices in their thermospheric density model
produces significant improvements in previous empirical
thermospheric density modeling.

\vskip12pt
\centerline {\bf2. Global activity indices}
\vskip12pt

\begin{figure}[h!]
 \centerline{\includegraphics[width=140mm]{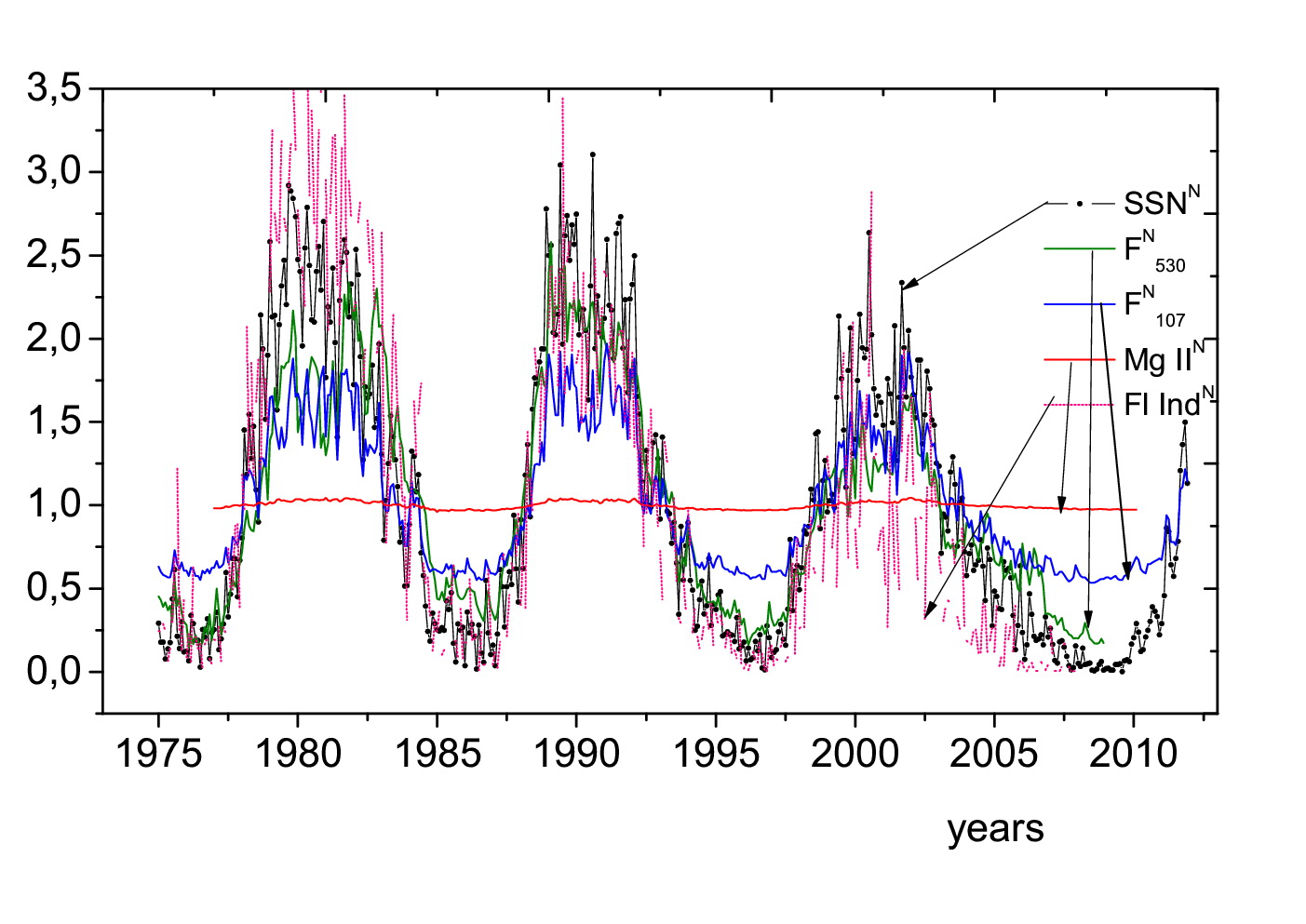}}
\caption{The time series of monthly average values of sunspot
numbers SSN, $F_{10.7}$, Mg II core-to-wing ratio, $F_{530}$, Flare
Index and Counts of flares. The upper index N indicates that solar
activity indices are normalized to their values averaged over the
analyzed time interval.} \label{Fi:Fig1}
\end{figure}
\medskip

Then we have to say a few words about solar indices studied in this
paper.

At Figure 1 we show the activity indices which are normalized to
their values averaged over the analyzed time interval. We can see in
most cases that the relative variation of an index in a solar cycle
is about 2-3 times. However, the magnitude of Mg II - index (as well
as TSI, not shown in this figure) changes very little - about shares
of percent.

The sunspot number SSN (also known as the International sunspot
number, relative sunspot number, or Wolf number) is a quantity that
measures the number of sunspots and groups of sunspots present on
the surface of the sun.

The historical sunspot record was first put by Wolf in 1850s and has
been continued later in the 20th century until today. Wolf's
original definition of the relative sunspot number for a given day
as $R = 10 \cdot$ Number of Groups + Number of Spots  visible on the
solar disk has stood the test of time. The factor of 10 has also
turned out to be a good choice as historically a group contained on
average ten spots. Almost all solar indices and solar wind
quantities show a close relationship with the SSN. (Svalgaard {\it
et al.} 2011; Svalgaard \& Cliver  2010). In our paper we use the
proper homogeneity calibrations of SSN from (National Geophysical
Data Center. Solar Data Service 2013).

At the present time the 10.7 - cm solar radio flux $F_{10.7}$ is
measured at the Dominion Radio Astrophysical observatory in
Penticton, British Columbia  by the Solar Radio Monitoring
Programme. $F_{10.7}$  is a useful proxy for the combination of
chromospheric, transition region, and coronal solar EUV emissions
modulated by bright solar active regions whose energies at the Earth
are deposited in the thermosphere.(Tobiska {\it et al.} 2008)
pointed the high EUV - $F_{10.7}$ correlation and used this in the
Earth's atmospheric density models.

According to Tapping \& DeTracey 1990 the 10.7 - cm emission from
the whole solar disc can be separated on the basis of characteristic
time-scales into 3 components: (i) transient events associated with
flare and similar activity having duration less than an hour; (ii)
slow variation in intensity over hours to years, following the
evolution of active regions in cyclic solar activity designated as
S-component; (iii) a minimum level below which the intensity never
falls - the "Quiet Sun Level". The excellent correlation of
S-component at 10.7 cm wavelength with full-disc flux in Ca II and
MgII was discussed by (Donnelly {\it et al.} 1983 ). The 10.7 cm
flux resembles the integrated fluxes in UV and EUV well enough to be
used as their proxy (Chapman  \& Neupert 1974; Donnelly {\it et
al.}1983; Bruevich \& Nusinov 1984; Nicolet \& Bossy 1985; Lean
1987)

This radio emission comes from high part of the chromosphere and low
part of the corona. $F_{10.7}$ radio flux  has two different
sources: thermal bremsstrahlung (due to electrons radiating when
changing direction by being deflected by other charged participles -
free-free radiation) and gyro-radiation (due to electrons radiating
when changing direction by gyrating around magnetic fields lines).
The (iii) a minimum level component (when SSN is equal to zero as it
was at the minimum of the cycle 24 and local magnetic fields are
negligible) is defined by free-free source. When the local magnetic
fields become strong enough at the beginning of the rise phase of
solar cycle and solar spots appear the gyro-radiation source of
$F_{10.7}$ radio flux begins to prevail over free-free so (i) and
(ii) components begin to grow strongly.

The S-component comprises the integrated emission from all sources
on solar disc. It contains contribution from free-free and
gyroresonance processes, and perhaps some non-thermal emission
(Gaizauscas \& Tapping 1998). The relative magnitude of these
processes is also a function of observing wavelength. Observations
of emission from active regions over the wavelength range 21-2 cm
suggest that at 21 cm, free-free emission is dominant, whereas at 6
cm, the contribution from gyroresonance is larger. At a wavelength
of 10 cm, the two processes are roughly equal in importance. At a
wavelength of 2-3 cm, the emission is again mainly free-free,
possible with a non-thermal component (Gaizauscas \& Tapping 1998).
The spatial distributions of two thermal processes are different;
the gyroresonant emission originates chiefly in the vicinity of
sunspots, where the magnetic fields are strong enough, while the
free-free emission is more widely-distributed over the host region
complex (Tapping \& DeTracey 1990).

The intensities of the Ca II and Mg II spectral lines are primary
functions of chromospheric density and temperature, while the soft
X-rays are produced in the corona. The high degree of correlation of
the 10.7 cm flux with all these quantities suggests some dependence
upon common plasma parameters and that their sources are spatially
close. Another strong correspondence is between 10.7 cm flux and
full-disc X-ray flux. When activity is high, they are
well-correlated; however, when activity is low, the X-rays are too
weak to be detected, while some 10.7 cm emission in excess of the
"Quiet Sun Level" is always present (Kruger 1979). Our study of the
connection between 10.7 cm flux and full-disc X-ray flux (Bruevich,
\& Yakunina 2011) also confirm the conclusions of (Kruger 1979).

 Thus we have enhanced 10.7 cm radiation when the
temperature, density and magnetic fields are enhanced. So $F_{10.7}$
is a good measure of general solar activity.

The green and red coronal lines observations was regularly started
from 1960 and the new solar index $F_{530}$ - the averaged intensity
of coronal flux at $530.3$ nm was introduced. We used NASA data from
several observatories and from satellite's instruments. These data
were modified to the common uniform system and are available as
archive data of (National Geophysical Data Center. Solar Data
Service 2013).

The 280 nm Mg II solar spectrum band contains photospheric continuum
and chromospheric line emissions. The Mg II {\it h} and {\it k}
lines at 279.56 and 280.27 nm, respectively, are chromospheric in
origin while the weakly varying wings or nearby continuum are
photospheric in origin. The instruments of the satellites observe
both features. The ratio of the Mg II variable core lines to the
nearly non-varying wings is calculated. The result is mostly a
measure of chromospheric solar active region emission that is
theoretically independent of instrument sensitivity change through
the time.

Mg II core-to wing ratio (crw) observations were made at NOAA series
operational satellites (NOAA-16-18), which host the Solar
Backscatter Ultra Violet (SBUV) spectrometer (Viereck, {\it et al.}
2001). This instrument can scatter solar Middle Ultra Violet (MUV)
radiation near 280 nm. The Mg II observation data were also obtained
from ENVISAT instruments. NOAA started in 1978 (during the
$21^{st}$, $22^{nd}$ and the first part of the $23^{rd}$ solar
activity cycles), ENVISAT was launched on 2002 (last part of the
$23^{th}$ solar activity cycle). Comparison of the NOAA and ENVISAT
MgII index observation data shows that both the Mg II indexes agree
to within about 0.5\% ( Puga \& Viereck 2004), (Scupin {\it et al.}
2005). We used both the NOAA and ENVISAT Mg II index observed data.

The Mg II index is especially good proxy for some Far Ultra Violet
(FUV) and Extreme Ultra Violet (EUV) emissions (Scupin {\it et al.}
2005). It well represents photospheric and lower chromospheric solar
FUV Schumann-Runge Continuum emission near 160 nm that maps into
lover thermosphere heating due to O2 photodissociation (Bowman, {\it
et al.}, 2008). Since a 160 nm solar index is not produced
operationally, The Mg II index proxy is used for comparison with the
other solar indices (Tobiska {\it et al.} 2008).

Solar irradiance is the total amount of solar energy at a given
wavelength received at the top of the earth's atmosphere per unit
time. When integrated over all wavelengths, this quantity is called
the total solar irradiance (TSI) previously known as the solar
constant. Regular monitoring of TSI has been carried out since 1978.
From 1985 the total solar irradiance was observed by Earth Radiation
Budget Satellite (EBRS). We use the NGDC TSI data set from combined
observational data of several satellites which were collected in
NASA archive data (National Geophysical Data Center. Solar Data
Service 2013). The importance of UV/EUV influence to TSI variability
(Active Sun/Quiet Sun) was pointed by (Krivova \& Solanki 2008).
There were indicated that up to 63.3\% of TSI variability is
produced at wavelengths below 400 nm. Towards activity maxima SSN
grows dramatically. But on average the Sun brightens about 0.1\%
only. This is due to the increase of amounts of bright and dark
features: faculae and network elements on the solar surface on the
one hand and spots on the other hand. The total area of the solar
surface covered by such features rises more strongly as the cycle
progresses than area of dark sunspots. The TSI (from Earth Radiation
Budget Satellite) maxima are fainter than those of the other indices
because the solar irradiance variation in solar cycle is
approximately equal to 0.14\%. This value seems very small but is
normal. Some TSI physics-based models have been developed with using
the combined proxies describing sunspot darkening (sunspot number or
areas) and facular brightening (facular areas, Ca II or Mg II
indices), see (Frontenla {\it et al.} 2004), (Krivova {\it et al.}
2003).

We also analyzed two activity indices which describe rapid processes
on the Sun - Flare Index (FI) and monthly Counts of grouped solar
flares (Count of flares). According to (Solar-Geophysical Data
Reports 2009) the term 'grouped' means observations of the same
event by different sites were lumped together and counted as one.

(Kleczek 1952) defined the value $FI = it $ to quantify the daily
flare activity over 24 hours per day. He assumed that relationship
roughly gave the total energy emitted by the flare and named it
flare index (FI). In this relation $i$ represents the intensity
scale of importance of the flare and and $t$ the duration of the
flare in minutes. In this issue we also used the monthly averaged FI
values.

So it should be noted that the data used in our article are not
uniform enough
 but we neglect this. We study
the behavior of solar indexes during cycles of activity as a whole.
Thus we analyze the general trend in their relationship.

\vskip12pt \centerline {\bf3. Recent changes in the Sun} \vskip12pt

The recent solar cycle 23 was the outstanding cycle for authentic
observed data from 1849 year. It lasted 12.7 years and was the
longest one for two hundred years of direct solar observations.
This cycle is the second component in the 22-year Hale magnetic
activity cycle but the $23^{rd}$ cycle was the first case of
modern direct observations (from 1849 to 2008) when Gnevyshev-Ol's
rule was violated: activity indices in cycle 23 had their maximum
values less then the values in the cycle 22 (but according to
Gnevyshev-Ol's rule the cycle 23 must dominate).

(Ishkov 2009) pointed that in this unusual cycle 23 the monthly
averages values for SSN during 8 months exceeded 113 and most of
sunspot groups were less in size, their magnetic fields were less
composite and characterized by the greater lifetime near $2^{nd}$
maximum in comparison with values near the $1^{st}$ maximum. SSN
reaches its first maximum 3.9 years after the beginning. After the
first maximum, the index decrease by 14\% (of that maximum). The two
maxima of this index have the same amplitude

(Nagovitsyn {\it et al.} 2012)  showed that for sunspot numbers SSN
we see the opposite long-scale trends (which influence to increase
or decrease of SSN) during over the last several solar cycles. In
(Nagovitsyn {\it et al.} 2012) it was analyzed the data set of SSN
from (Penn \& Livingston 2006, Pevtsov {\it et al.} 2011) and it was
shown that during the cycle 23 and the beginning of the cycle 24 the
number of large sunspots gradually decreased, while the number of
small sunspots steadily increased. It was suggested that this change
in the fraction of small and large sunspots (perhaps, due to changes
in the solar dynamo) can explain the gradual decline in average
sunspot field strength as observed by (Penn \& Livingston 2006).

In the cycle 23 the $F_{10.7}$ radio flux and the TSI have the
lowest values from 2007 to 2009 (the beginning of the cycle 24) all
over of these indices observation period. The $F_{10.7}$ radio flux
index shows the second maximum is 8.4\% stronger than the first one.

It's known that all solar indices have been closely correlated as
they all derive from the same source: the variable magnetic field.
But while there has long been a stable relationship between the 10.7
cm flux and SSN this relationship has steadily deteriorated in the
past decade to the point where the sunspot number for a given flux
has decreased by about a third.

Observations by Livingston since 1998 until the present show that
the average magnetic field in sunspots has steadily decreased by
25\% (Livingston {\it et al.}, 2012). Since their magnetic fields
cool sunspots, a decreasing field means that sunspots are getting
warmer and that their contrast with the surrounding photosphere is
getting smaller, making the spots harder to see. Without the dark
spots, TSI might even be a bit higher, see (Svalgaard {\it et al.}
2011). We can see that the relation TSI/$F_{10.7}$ is lager a little
for the cycle 23 (Fig. 3c) compared to the previous cycle 22. It is
not clear what this will mean for the impact of solar activity on
the Earth's environment.

(Janardhan {\it et al.} 2010) have examined polar magnetic fields
for the last three solar cycles 21, 22 and 23 using NSO Kitt Peak
synoptic magnetograms and showed a large and unusual drop in the
absolute value of the polar fields during cycle 23 compared to
previous cycles and also it's association with similar behavior in
meridional flow speed.

\begin{figure}[h!]
   \centerline{\hspace*{0.005\textwidth}
               \includegraphics[width=70mm]{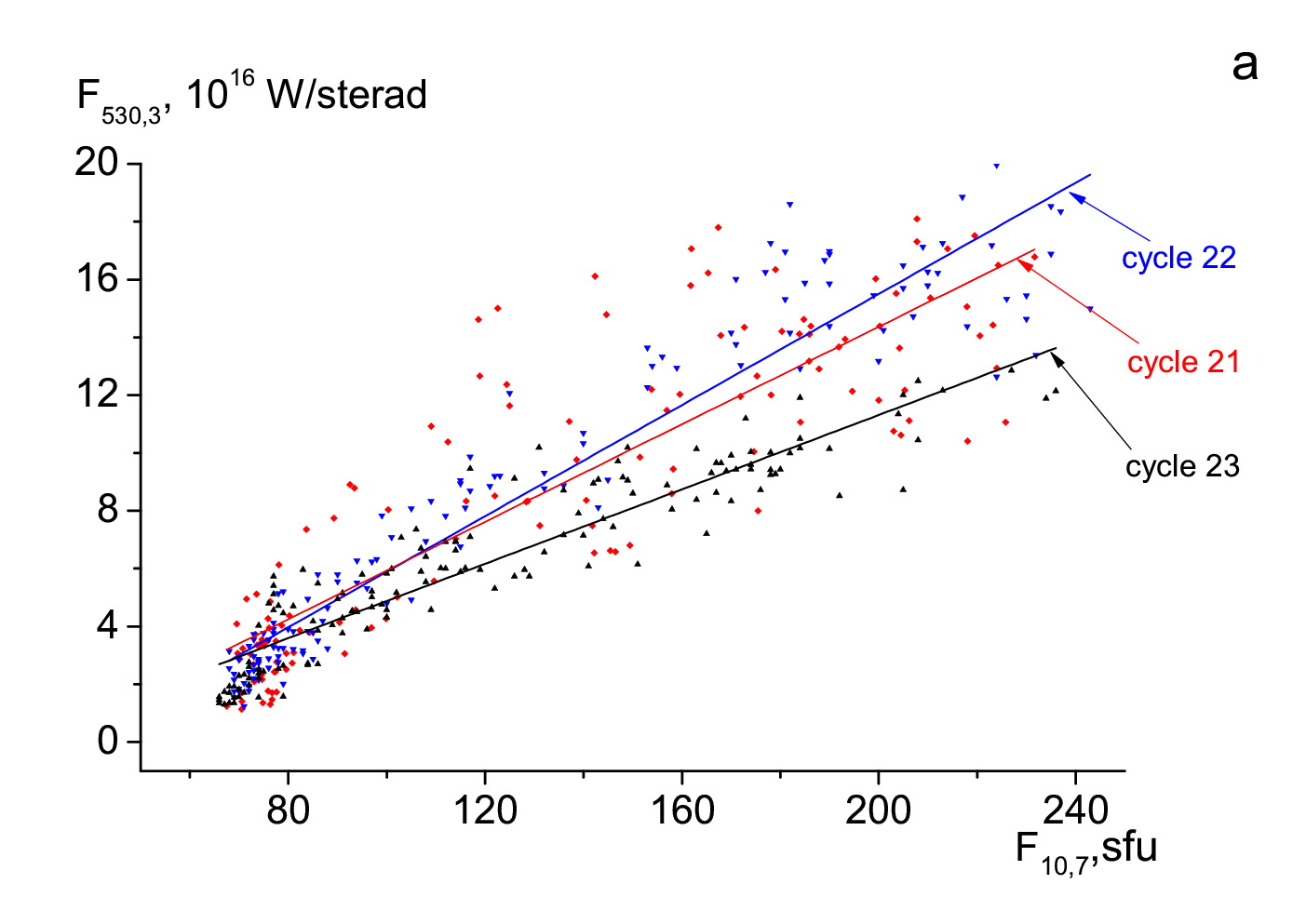}
               \includegraphics[width=70mm]{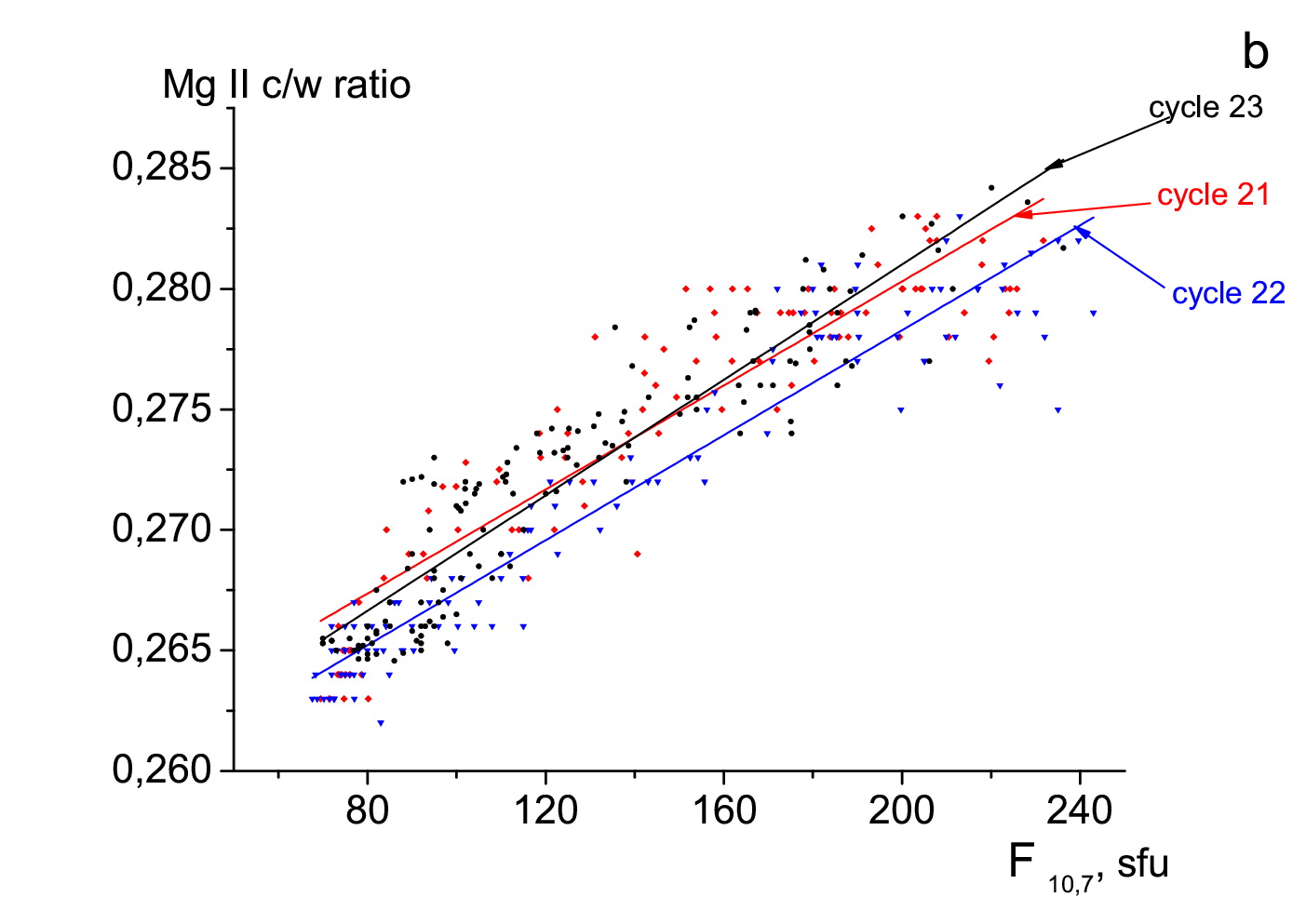}
              }
   \centerline{\hspace*{0.015\textwidth}
               \includegraphics[width=70mm]{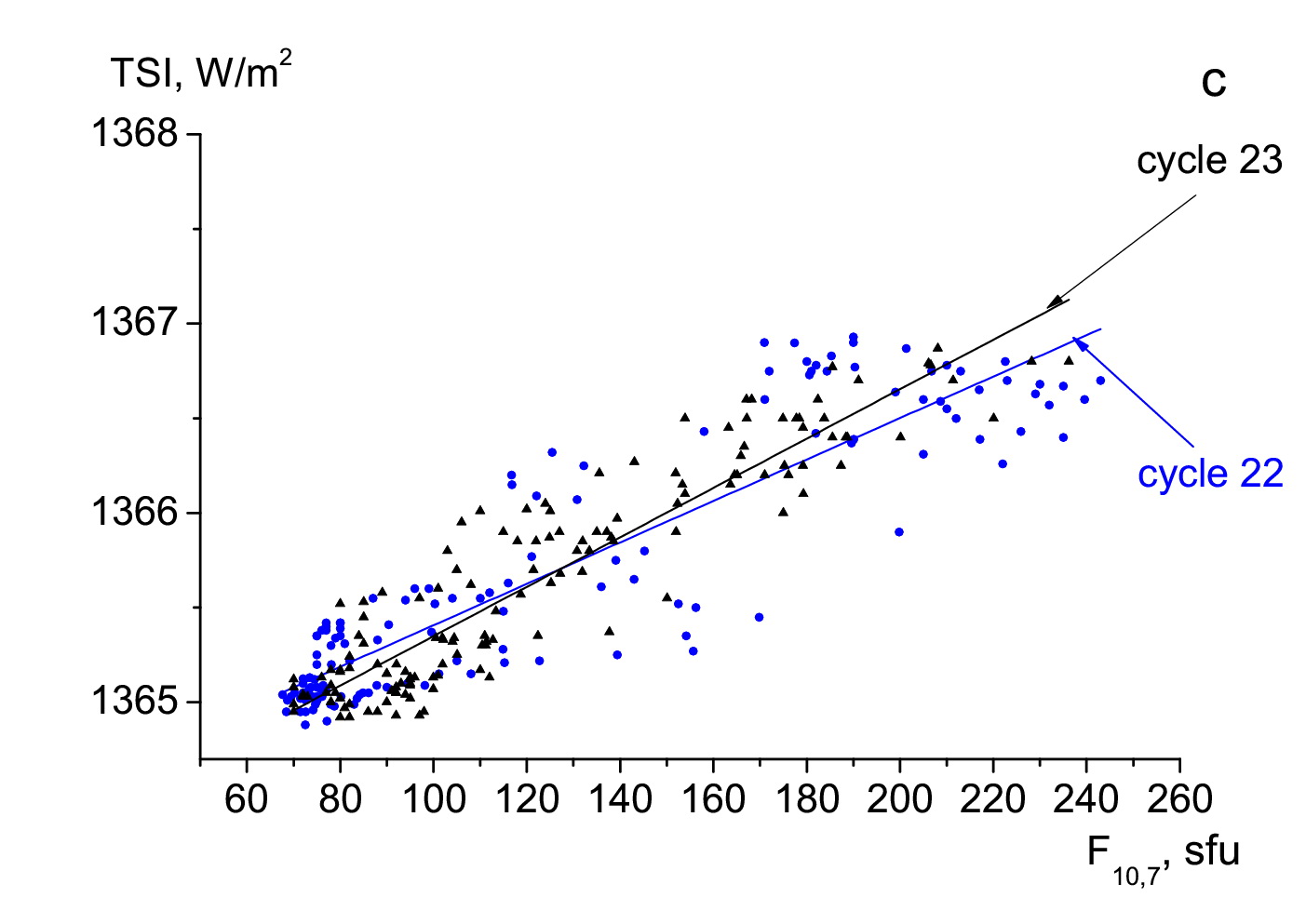}
               \includegraphics[width=70mm]{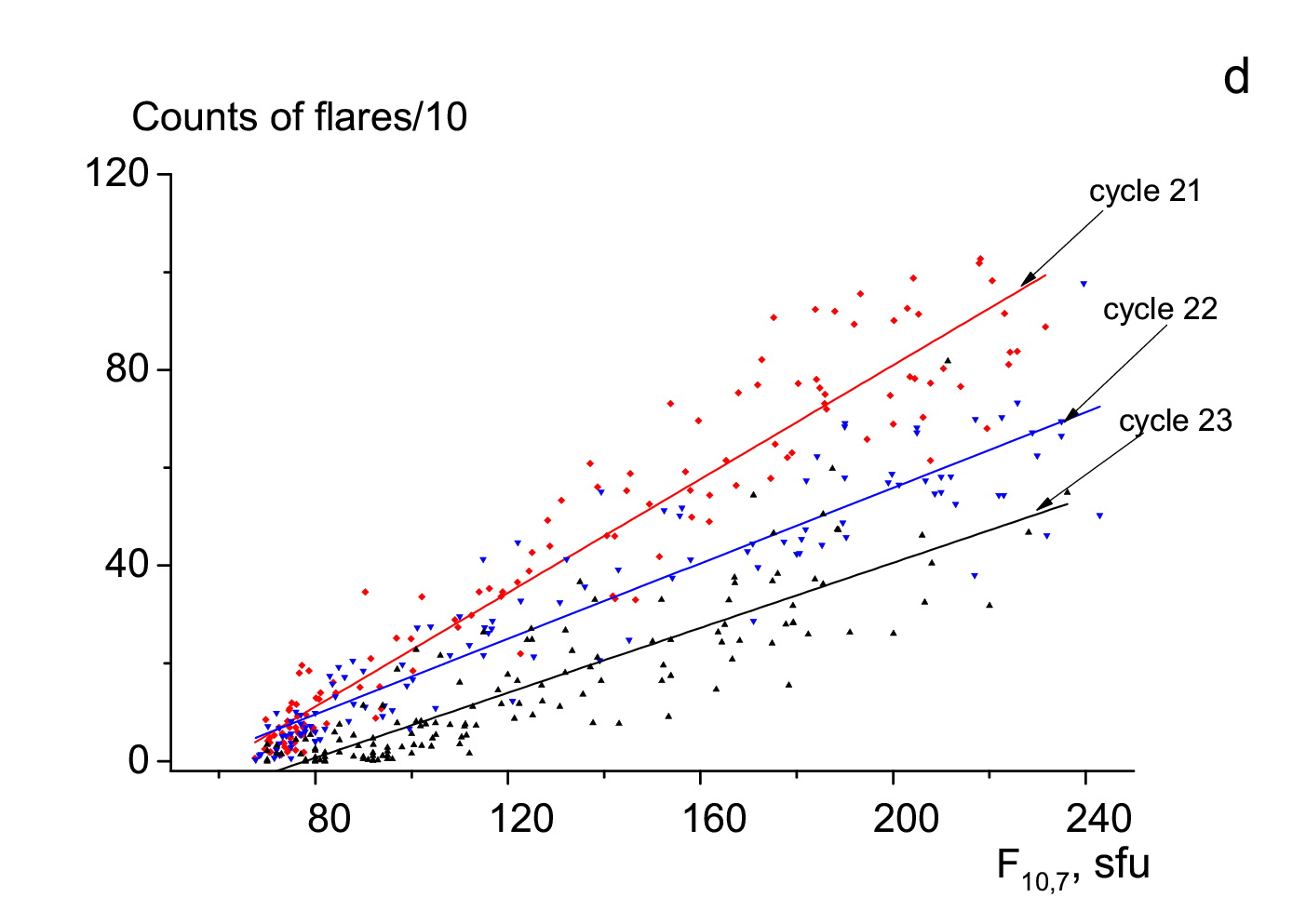}
              }

\caption{Correlation between monthly averages for solar indices
versus $F_{10.7}$ individually determined for the cycles 21, 22
and 23. (a) $F_{530}$, (b) Mg II core-to-wing ratio, (c) TSI and
(d) Counts of flares/10.
        } \label{Fi:Fig2}
   \end{figure}

In the cycle 23 the Flare index has a higher first maximum. This
shows that the flares can be more efficiently generated during the
first maximum, and it seems that the generation is decreasing
towards the end of the cycle.

 Figure 1 demonstrates that for all activity indices in the
$23^{rd}$ solar activity cycle we can see two maximums separated
one from another on 1.5 year approximately. We see the similar
double-peak structure in cycle 22 but for the cycle 21 the
double-peak structure is not evident. We see that there are
displacements in both maximum occurrence time of all these indices
in the $23^{rd}$ solar cycle.

Figure 1 also shows that for all solar indices in the cycle 23 the
relative depth of the cavity between two maximums is about
$10-15\%$.

Ishkov (2009) pointed that there was very high level of flared
activity in the cycle 21 and very low level of flared activity in
the cycle 23. The Sun's flare activity is an important indicator of
the general level of activity of the atmosphere is also described in
other activity indices, in particular around the solar disk index
and a locally varying flux in the H-alpha, see Bruevich (1995).

\vskip12pt
\centerline {\bf4. Changed relation between $F_{10.7}$
and solar activity indices in the cycles 21-23}
\vskip12pt

Figure 2 illustrates the high level of interconnection between
solar activity indices versus $F_{10.7}$. We see that coefficients
of linear regression (slope - $A$ and intercept - $B$) differ
among themselves for the activity cycles 21 - 23. The most
differences shows the Counts of flares index.

When studied solar activity indices in 21, 22 and 23 solar activity
cycles we separated out rise phases, cycle's maximum phase, cycle's
minimum phase and decline phase. In case of linear regression we've
found that the maximum values of correlation coefficients $K_{corr}$
reached for the rise and decline phases of the cycles. According to
our calculations the highest values of correlation coefficients
$K_{corr}$ we see in connection between SSN and $F_{10.7}$.
 Correlation coefficients
$K_{corr}$ for linear regression for TSI versus $F_{10.7}$ are the
minimal of all correlation coefficients determined here.

 The cyclic variation of fluxes in
different spectral ranges and lines at the 11-year time scale
 are widely spread phenomenon for F, G and K stars
(not only for the Sun), see (Bruevich \& Kononovich 2011, Bruevich
\& Ivanov-Kholodnyj 2011). The chromospheric activity indices
(radiative fluxes at the centers of the $H$ and $K$ emission lines
of Ca II - $396.8$ and $393.4$ nm respectively) for solar-type stars
were studied during HK-project by (Baliunas {\it et al.} 1995) at
Mount Wilson observational program during 45 years, from 1965 to the
present time. Authors of the HK-project supposed that all the
solar-type stars with well determined cyclic activity about 25\% of
the time remain in the Maunder minimum conditions. Some scientists
proposed that the solar activity in the cycle 24 will be very low
and similar to activity during the Maunder minimum period. We can
see now that although the new cycle of activity is not characterized
by very high SSN values, however, the activity in the cycle 24 is
significantly higher than in the Maunder minimum period.

\begin{figure}[h!]
   \centerline{\hspace*{0.0005\textwidth}
               \includegraphics[width=70mm]{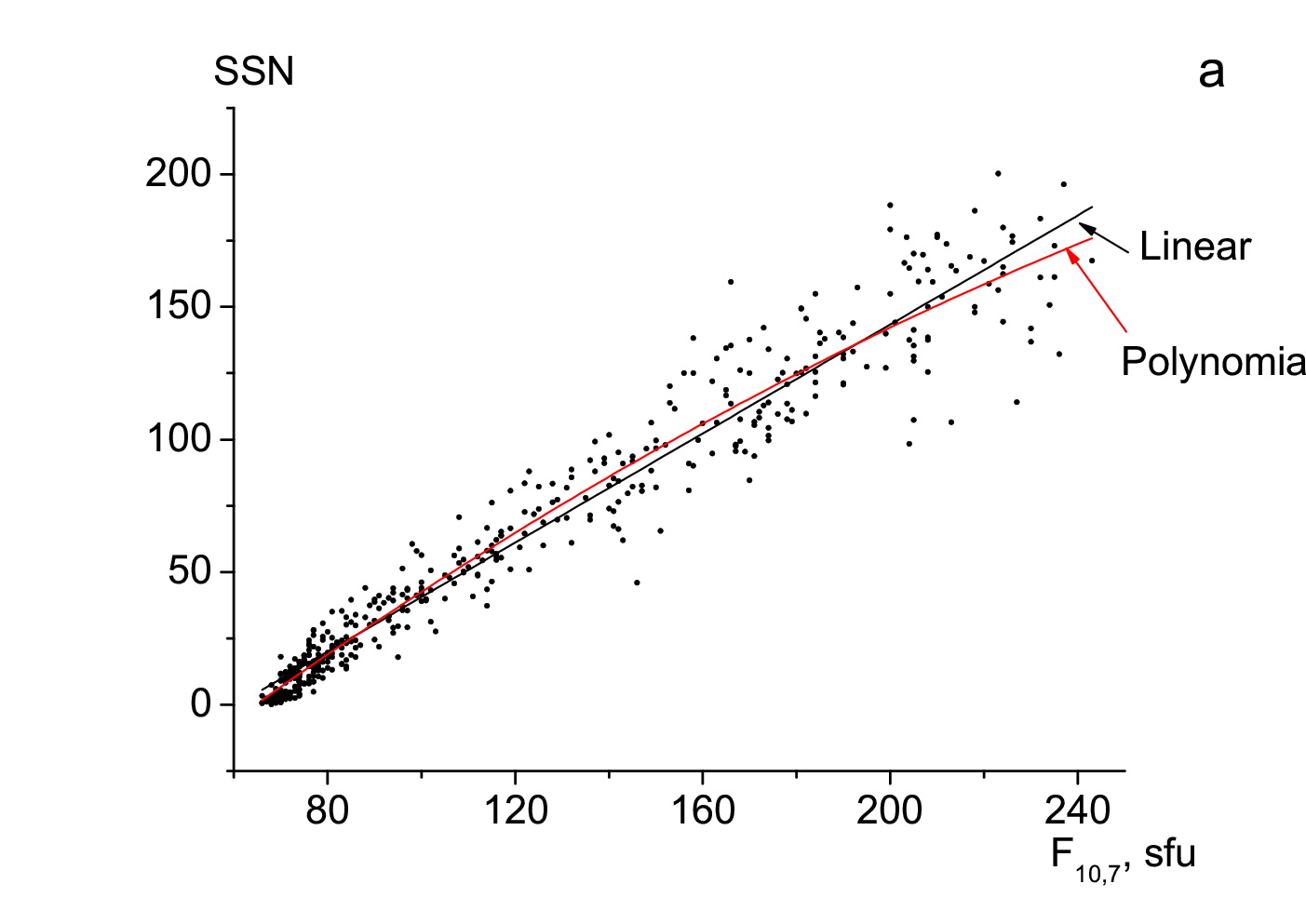}
               \includegraphics[width=70mm]{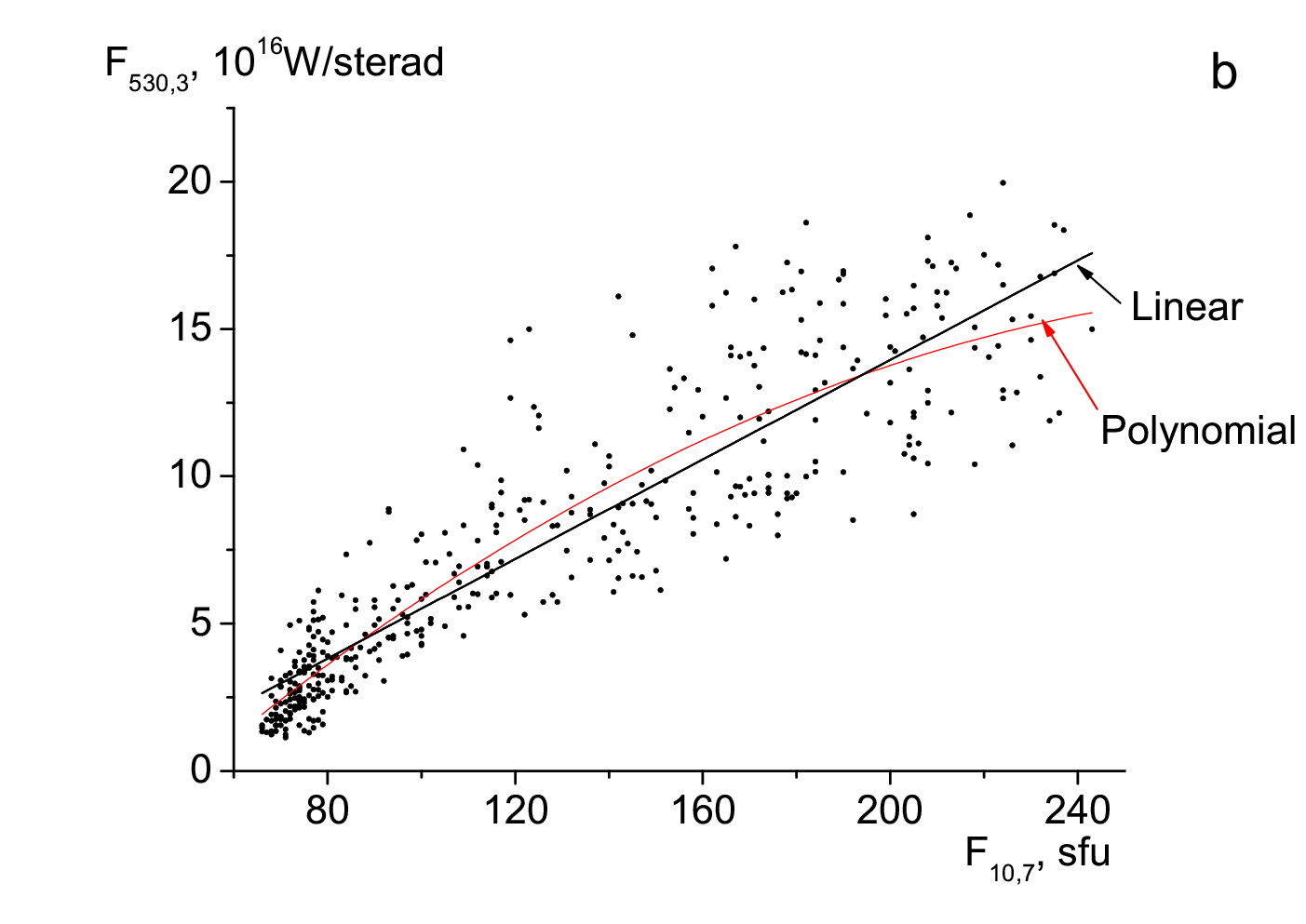}
              }
   \centerline{\hspace*{0.015\textwidth}
               \includegraphics[width=70mm]{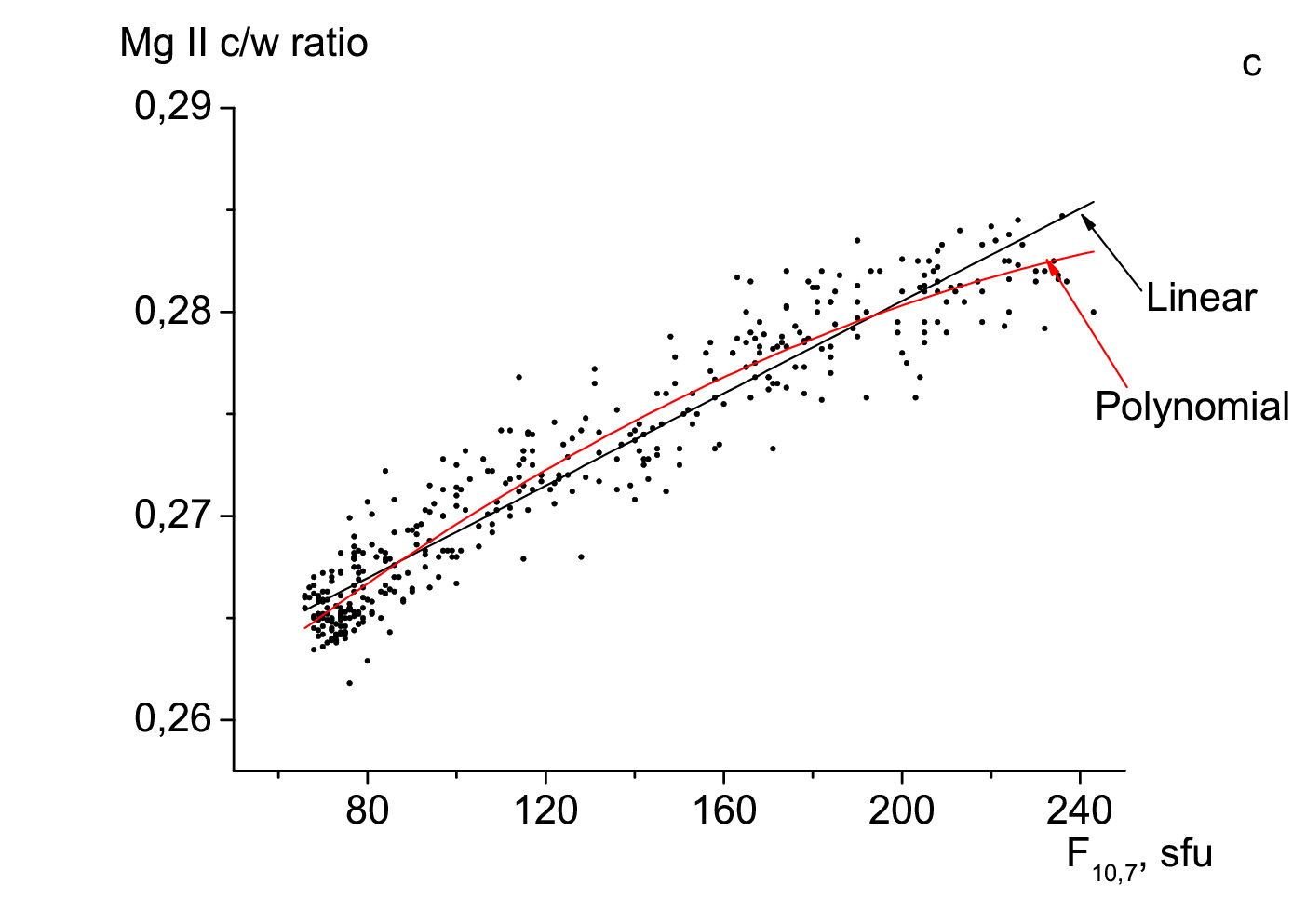}
               \includegraphics[width=70mm]{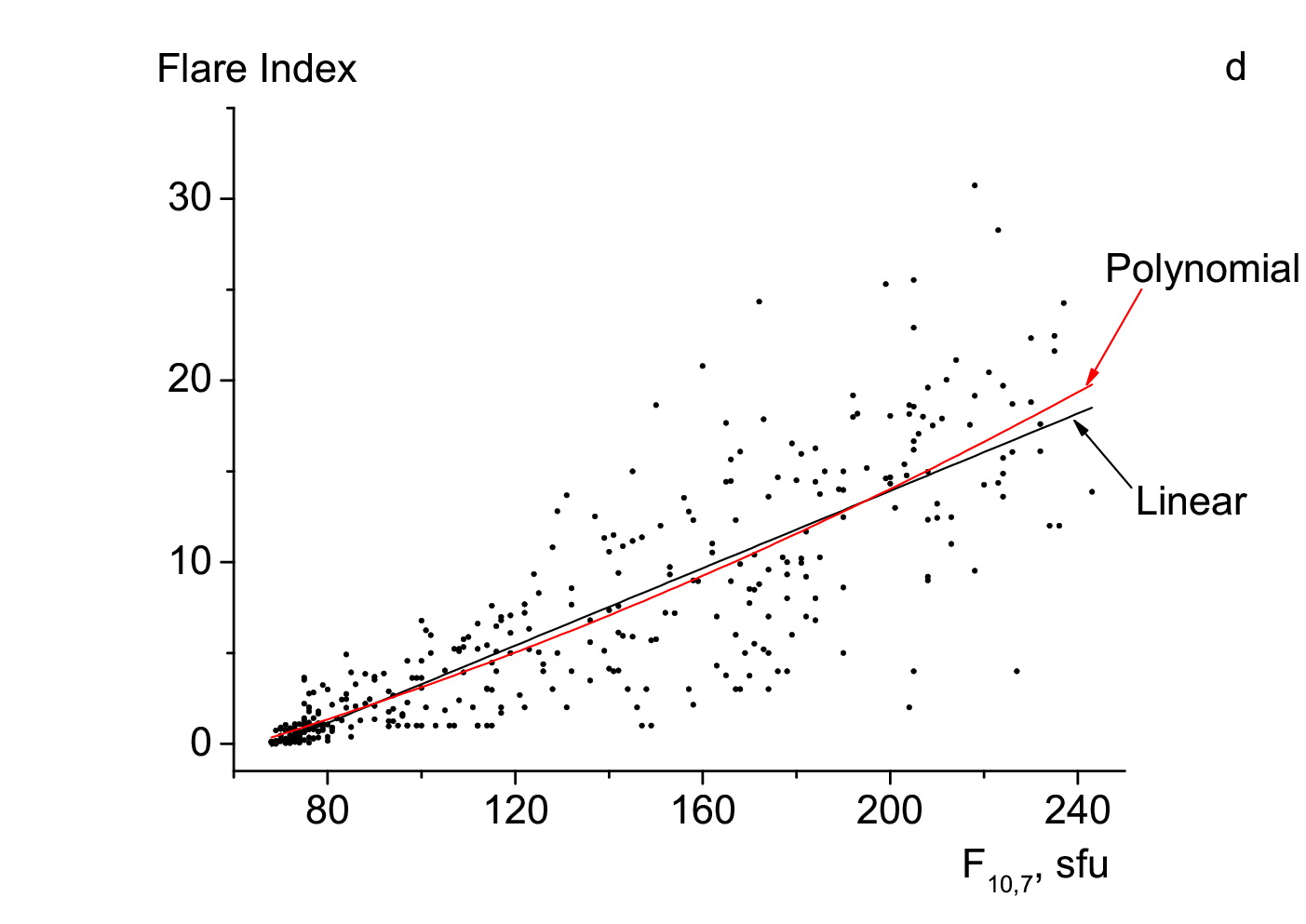}
              }
   \centerline{\hspace*{0.015\textwidth}
               \includegraphics[width=70mm]{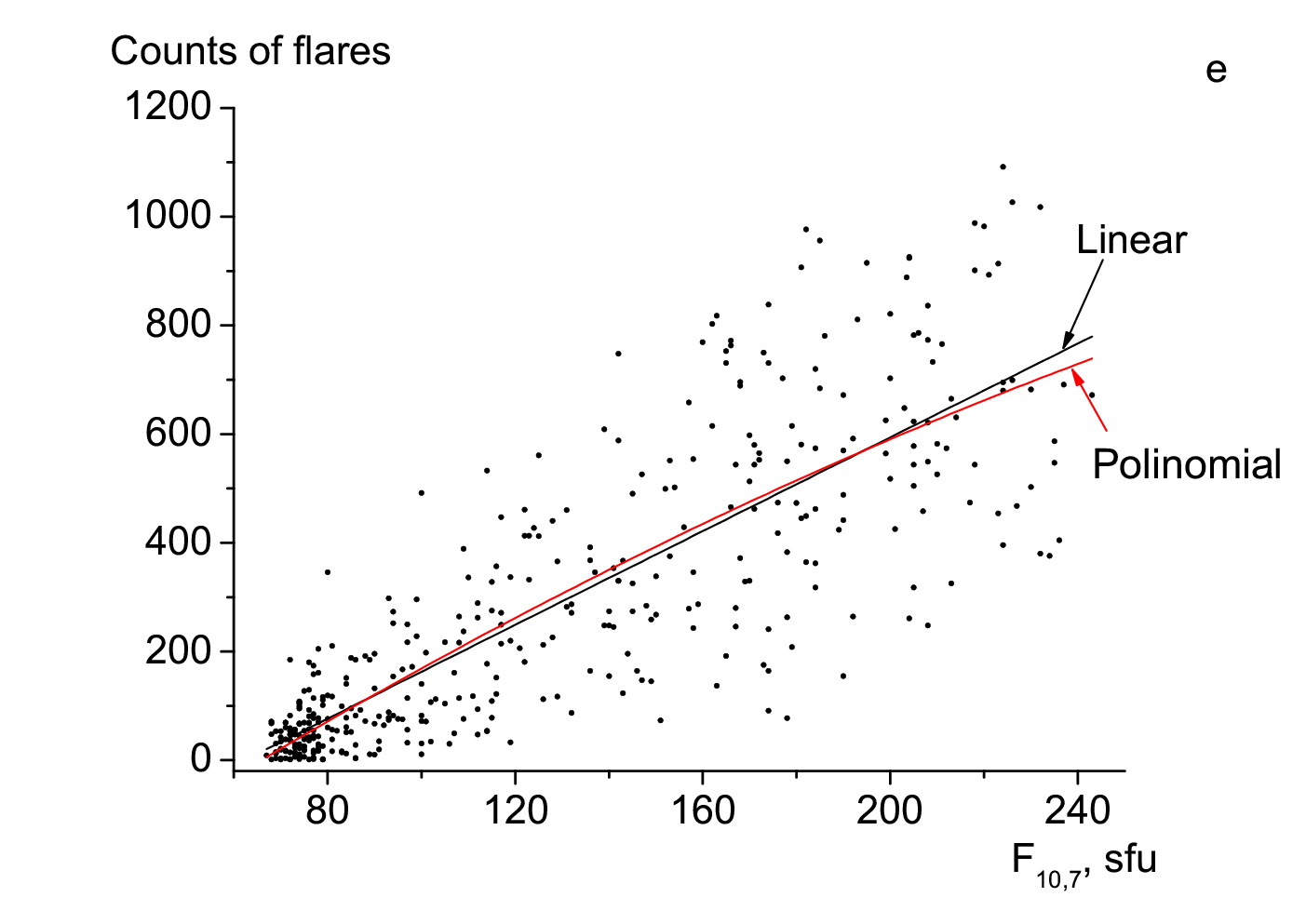}
               \includegraphics[width=70mm]{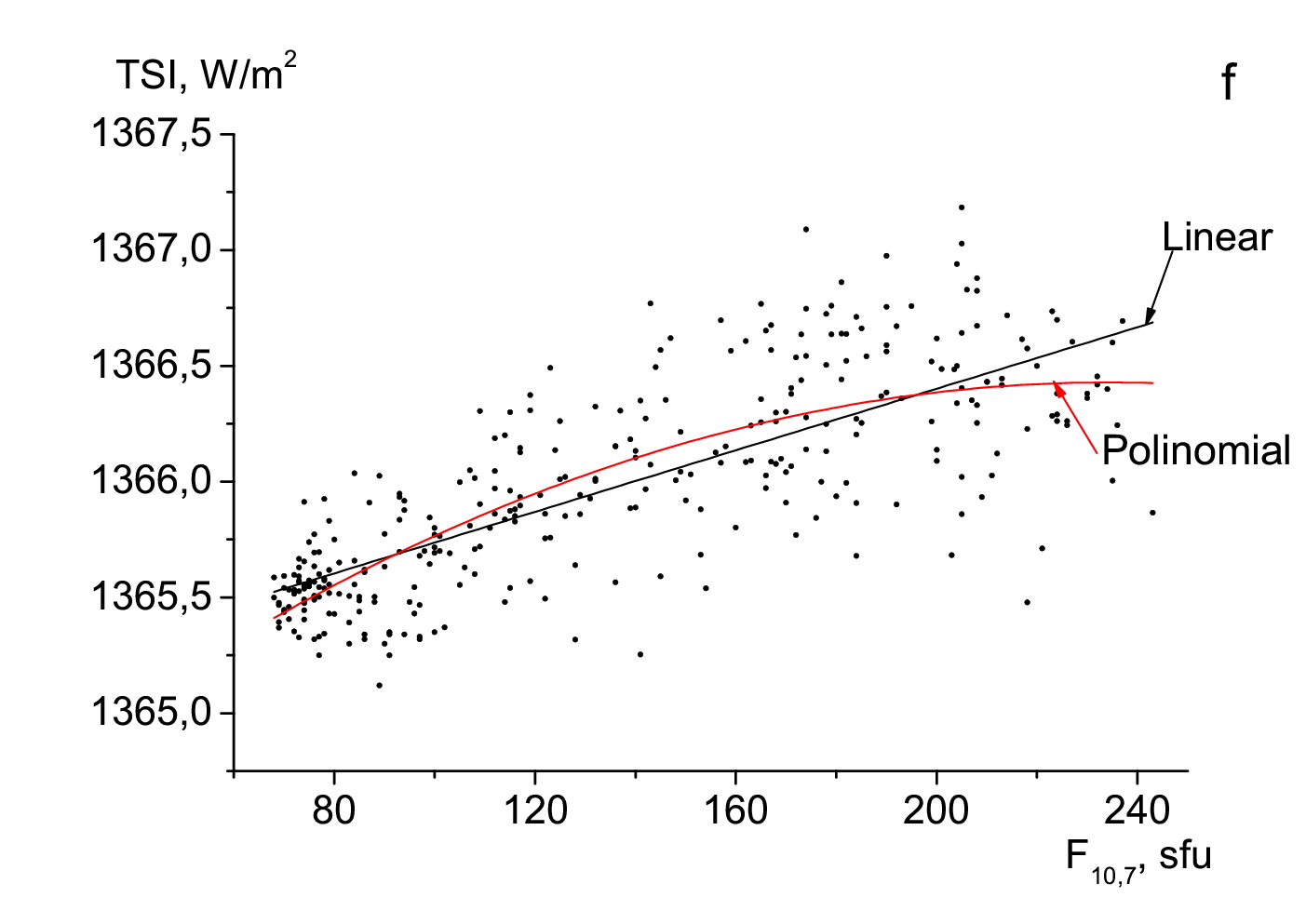}
               }

\caption{Correlation between monthly averages of solar indices
versus $F_{10.7}$ in the cycles 21 - 23 (1975 - 2008). (a) SSN,
(b) $F_{530}$, (c) Mg II core-to-wing ratio, (d) Flare Index, (e)
Counts of flares and (f) TSI.} \label{Fi:Fig3}

   \label{F-6panels}
   \end{figure}

 At Figure 3 we presented the interconnection
between solar indices and radio flux $F_{10.7}$ in the cycles 21 -
23. We analyzed the interconnection between activity indices SSN,
$F_{530}$, Mg II core to wing ratio, Flare Index, Counts of flares
and TSI versus $F_{10.7}$  for the $21^{st}$, $22^{nd}$ and
$23^{th}$ solar cycles. We have studied both the linear and
polynomial dependencies.

The linear model corresponds to the linear regression equation:

\begin{equation}
    F_{ind}  = A_{ind} + B_{ind} \cdot F_{10.7} \,.
   \end{equation}

were  $F_{ind}$ is the activity index flux,

$A_{ind}$ is the intercept of a linear regression,

$B_{ind}$ is the slope of a linear regression.

\begin{table}[h!h!h!]

\begin{minipage}[]{140mm}

\small \centering

\caption{Solar activity indices versus $F_{10.7}$. Coefficients of
linear regressions: $A$, $B$ and their standard errors.
Observational data 1975 - 2010. } \label{T-simple}

\end{minipage}

\small \centering

\begin{tabular}{lllll}   
\hline\noalign{\smallskip}
 Act. indices        & $A_{ind}$        & $B_{ind}$       &  Err. $\sigma_A$&  Err. $\sigma_B$ \\
 versus $F_{10.7}$&          &       &       &     \\
   \hline\noalign{\smallskip}
 SSN        & -62.28     & 1.03    & 1.57                  &   0.011                         \\
 $F_{530}$  & -2.93      & 0.084   & 0.27                  &   0.002                         \\
 Mg II         & 0.258  & $1.3 \cdot 10^{-4}$  & $2.6 \cdot 10^{-4}$  &  $1.8 \cdot 10^{-7}$                   \\
 Flare Index   & -7.37      & 0.106      &  0.51               &    0.0036                  \\
 Counts fl/10  & -269.03  & 4.31     &    20.14              &   0.146                      \\
 TSI         & 1365.07    & 0.0066      &  0.044               &  $3 \cdot 10^{-4}$                  \\
\hline\noalign{\smallskip}

\end{tabular}

\end{table}

 In the Table 1 we present the coefficients of the linear regressions
 and their standard errors $\sigma$ of intercept and slope values.

The polynomial model corresponds to the following equation of a
second order polynomial:

\begin{equation}
    F_{ind}  = A_{ind} + B1_{ind} \cdot F_{10.7} + B2_{ind} \cdot F^2_ {10.7} \,.
   \end{equation}

were  $F_{ind}$ is the activity index flux,

$A_{ind}$ is the intercept of a polynomial regression,

$B1_{ind}$ and $B2_{ind}$ are the coefficients of a polynomial
regression.

\begin{table}[h!h!h!]

\small \centering

\begin{minipage}[]{140mm}

\caption{Solar activity indices versus $F_{10.7}$. Coefficients of
polynomial regressions: $A$, $B1$, $B2$ and their standard errors.
Observational data 1975 - 2010.}
\label{}
\end{minipage}

\tabcolsep 3mm

 \begin{tabular}{llllllll}     
  \hline\noalign{\smallskip}
 Act. indices & A      &  B1   & B2    &Err. $\sigma_A$&Err. $\sigma_{B1}$&Err. $\sigma_{B2}$\\
 versus $F_{10.7}$ &          &       &       &                 &
            &      \\
   \hline\noalign{\smallskip}

 SSN  & -87.26           & 1.45  & -0.0015  & 4.86  & 0.078 &  $3 \cdot 10^{-4}$  \\
 $F_{530}$ & -7.38    & 0.158 & $-2,6 \cdot 10^{-4}$  & 0.84 & 0.013 &  $4.8 \cdot 10^{-5}$  \\
 Mg II      & 0.25     & $2 \cdot 10^{-4}$  & $-3 \cdot 10^{-7}$  & $7 \cdot 10^{-4}$ & $1 \cdot 10^{-5}$ &  $4 \cdot 10^{-8}$      \\
 Flare Index  & -4.36   & 0.057   &  $1.7 \cdot 10^{-4}$  & 1.58 & 0.024 &  $7 \cdot 10^{-5}$  \\
 Counts fl/10 & -361.4 & 5.83   & -0.005  & 64.01  & 1.01 & 0.0035            \\
 TSI          & 1364.4   & 0.017  & $-3,7 \cdot 10^{-5}$ & 0.13 & 0.002 & $7 \cdot 10^{-7}$ \\
  \noalign{\smallskip}\hline

\end{tabular}
\end{table}

 In the Table 2 we present the coefficients of
polynomial regressions: $A$, $B1$, $B2$ and their standard errors
$\sigma$ of intercept and slope values.

We have to point out that close interconnection between radiation
fluxes characterized the energy release from different atmosphere's
layers is the widespread phenomenon among the stars of late-type
spectral classes. (Bruevich \& Alekseev 2007) confirmed that there
exists the close interconnection between photospheric and coronal
fluxes variations for solar-type stars of F, G, K and M spectral
classes with widely varying activity of their atmospheres. It was
shown that the summary areas of spots and values of X-ray fluxes
increase gradually from the sun and HK-project stars with the low
spotted discs to the highly spotted K and M-stars for which
(Alekseev \& Gershberg 1996) constructed the zonal model of the
spots distributed at the star's disks. The variations of activity
indices in the whole 11-yr cycle of the Sun are very similar to the
cyclical variations of the chromospheric fluxes on the stars. So we
can simulate the dependencies which describe the variations of the
indices during the activity cycle for the stars as for the Sun, see
Bruevich and Bruevich (2004).

\vskip12pt \centerline {\bf5. The time variations of correlation
coefficient $K_{corr}(t)$ for the linear} \centerline {\bf
regression of solar activity indices versus $F_{10.7}$ and versus
SSN. } \vskip12pt

\begin{figure}[h!]
   \centerline{\hspace*{0.015\textwidth}
               \includegraphics[width=70mm]{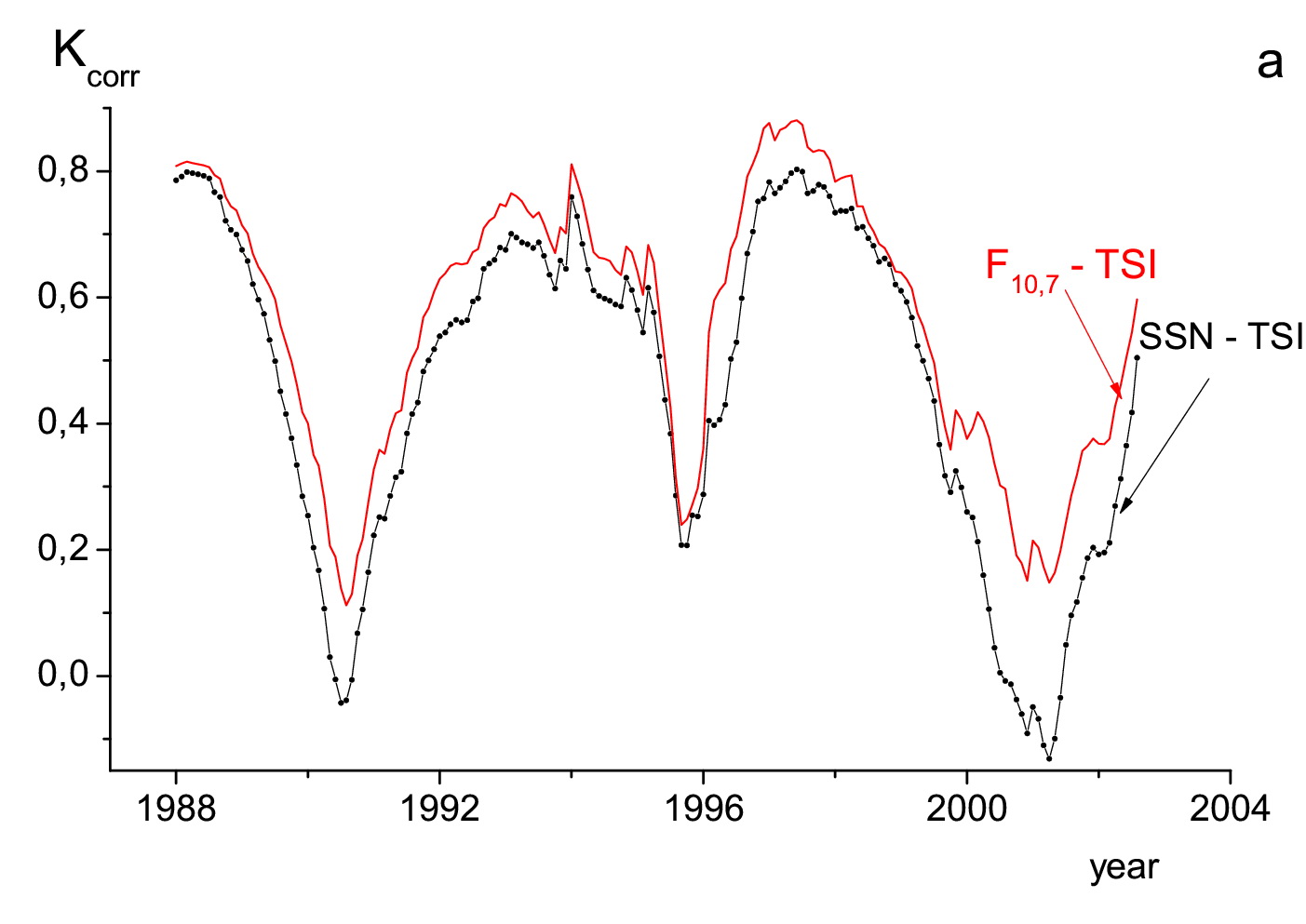}
                \includegraphics[width=70mm]{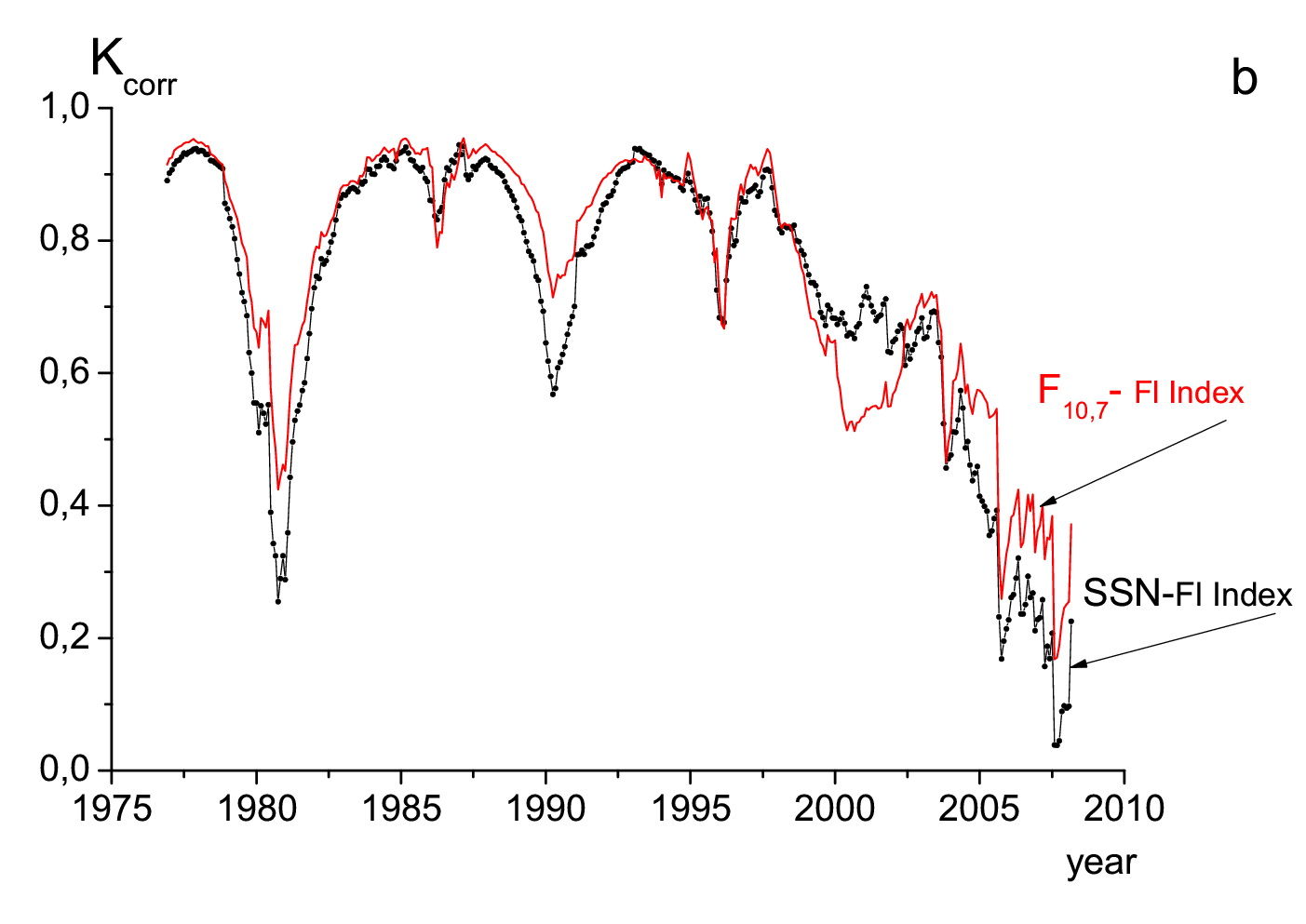}
              }
   \centerline{\hspace*{0.015\textwidth}
               \includegraphics[width=70mm]{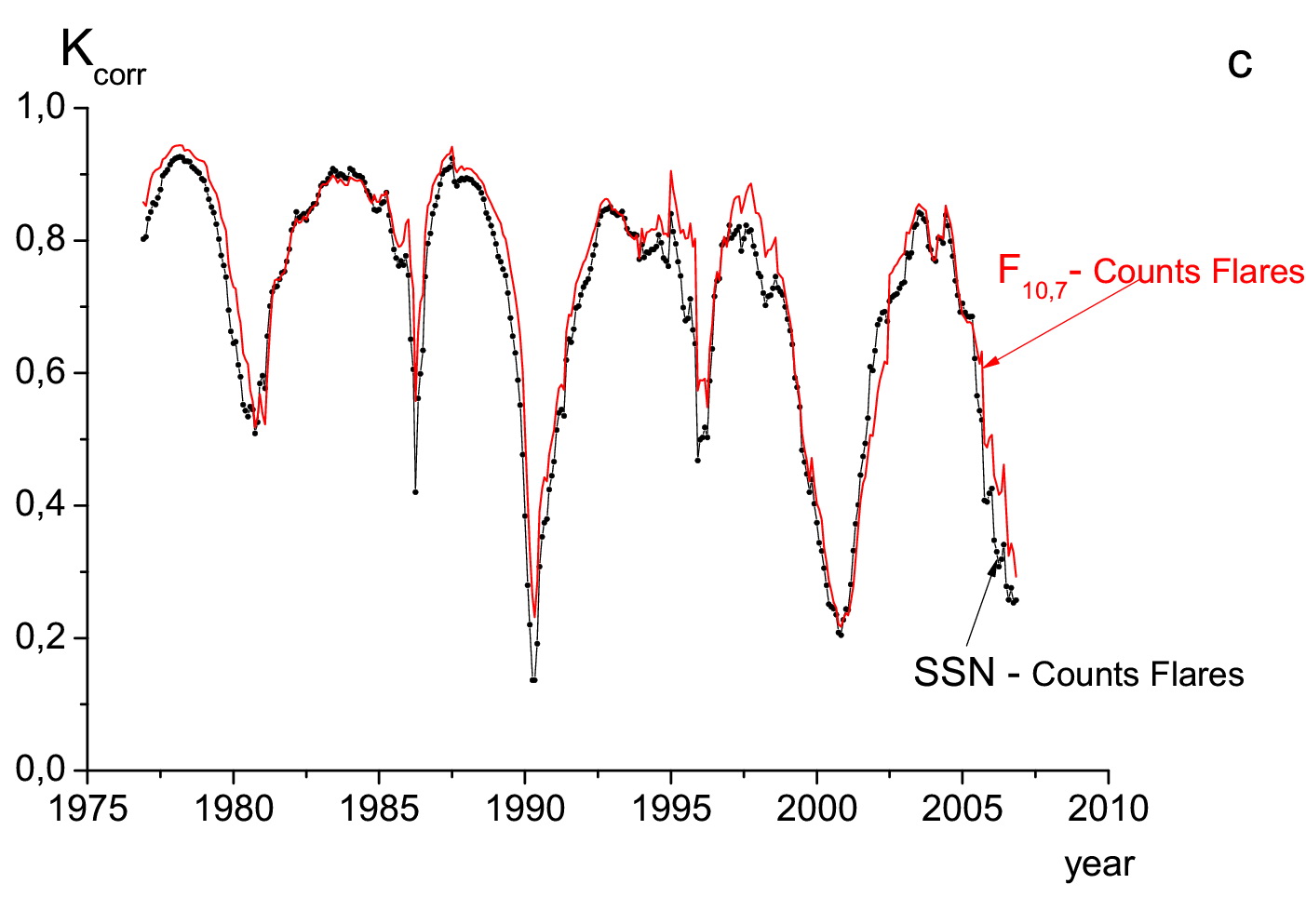}
                \includegraphics[width=70mm]{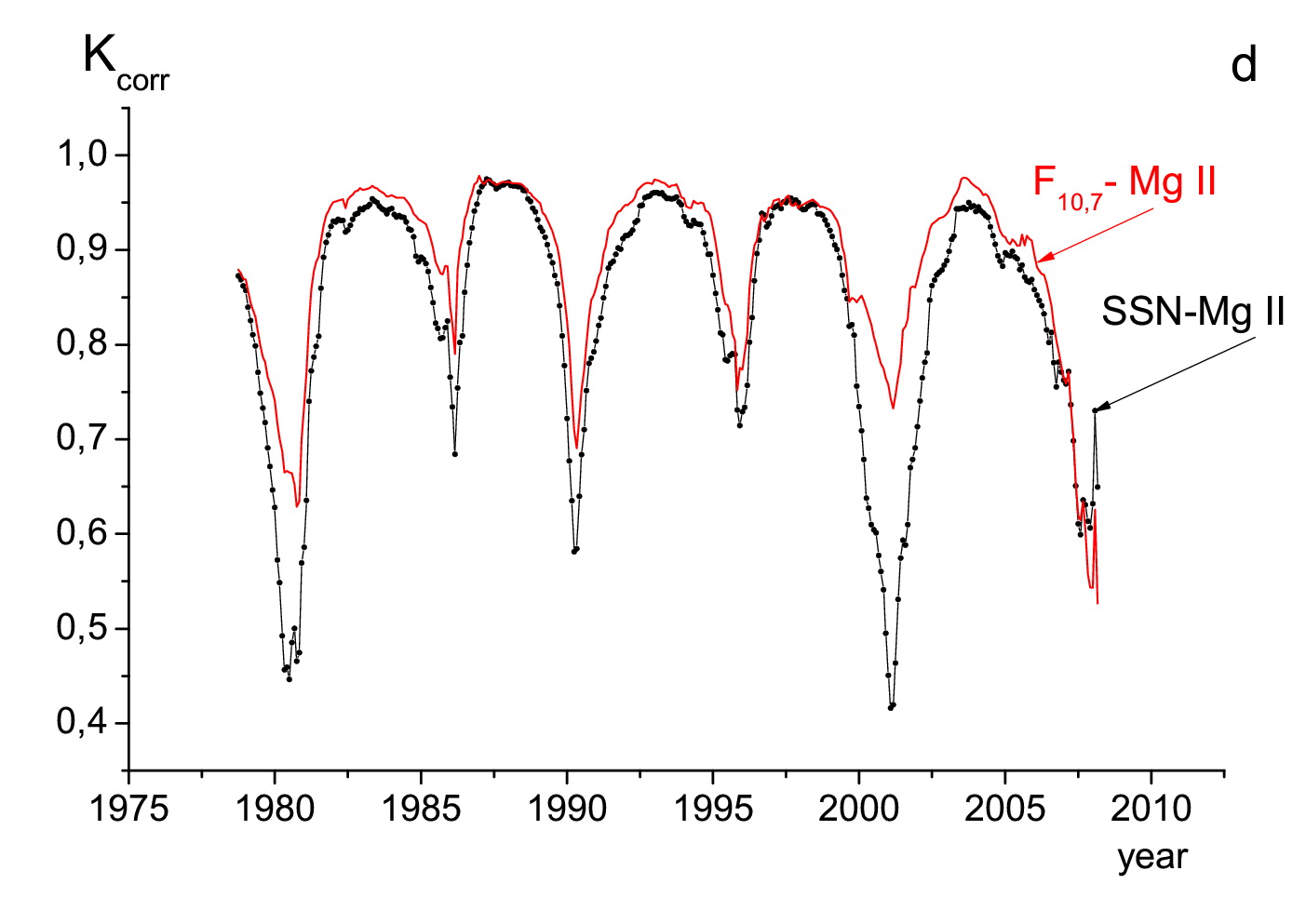}
              }

\caption{Correlation coefficients of linear regression
$K_{corr}(t)$ for (a) TSI, (b) Flare Index, (c) Counts of flares
and (d) Mg II UV-index versus SSN and versus $F_{10.7}$.
$K_{corr}(t)$ was calculated during three-year interval.
Observational data 1975 - 2010).
        } \label{Fi:Fig4}
   \label{F-4panels}
   \end{figure}

We've calculated values $K_{corr}(t)$ of linear regression for solar
activity indices versus $F_{10.7}$ for the cycles 21,22 and 23. The
values $K_{corr}(t)$ were determined for each moment of time $t$
from $K_{corr}(t)$ calculation during the 3-year time interval
$~t-1.5yr~< \Delta T<~t+1.5 yr $.

Figure 4 demonstrates the results of our correlation calculations
of these solar activity indices versus $F_{10.7}$ and versus SSN
 - $K_{corr}(t)$ variations during the cycles 21 - 23. We can see
that all the
  $K_{corr}(t)$ values have the maximum amounts at the rise and at
the decline phases. The minimum values of the $K_{corr}(t)$ we see
at the minimum and the maximum phases of solar cycles.

We can see that the minimum values of
 correlation coefficients $K_{corr}(t)$ for the solar
indices versus $F_{10.7}$ and SSN occurred twice during the 11-year
cycle We assumed that this fact
 must be considered for the understanding of the solar indices interconnections and
 for successful forecasts of different
 activity indices using $F_{10.7}$ or SSN observations.

Note that the linear correlation (see Figure 4) of activity
indices $F_{530}$ , MgII index, Flare Index and TSI versus
$F_{10.7}$ is a little stronger  than the linear correlation of
these indices versus SSN. We assumed that it’s a logical result:
these indices (as well $F_{10.7}$) characterize the solar
irradiance proceeded from different altitudes of solar atmosphere.
But SSN and Counts of flares are not connected directly with solar
irradiance at different wavelengths or spectral intervals. So the
linear correlation  of Counts of flares (see Figure 4) has no
difference versus $F_{10.7}$ or versus SSN because the Counts of
flares index describes the fast flared processes (not irradiance)
on the Sun while SSN is the relatively subjective measure of the
total level of solar activity

The cyclic behavior of $K_{corr}$ can be explained by following
assumption: we imagine that some activity index flux depends on time
$t$ by the expression:

\begin{equation}
    F_{ind}(t) = F_{ind}^{background}(t) + \Delta F_{ind}^{AR}(t) \,.
   \end{equation}

were $F_{ind}^{background}(t)$ is the background flux which
continuously rising with increasing of solar activity. Background
flux consists of two components - (1) slow variation in intensity
over hours to years, following the evolution of active regions in
cyclic solar activity and (2) a minimum level below which the
intensity never falls - the "Quiet Sun Level". In case of the radio
flux $F_{10,7}$ the component (1) is (ii) , the component (2) is
(iii) respectively. It can be note that all indices have the  "Quiet
Sun Level" different from zero except SSN, which at the minimum of
the cycle has the "Quiet Sun Level" value equal to zero.

$\Delta F_{ind}^{AR}(t)$ is the additional flux to the overall flux
from the active regions.

The previous correlation study allows us to consider that
$F_{ind}^{background}(t)$ and $\Delta F_{ind}^{AR}(t)$ are the
linear functions of the background and activity regions levels of
solar activity. In our case we choose the radio flux $F_{10,7}$ as
the best basal indicator of solar activity levels:

\begin{equation}
    F_{ind}^{background}(t) = a_1 + b_1 \cdot
F_{10.7}^{background}(t) \,.
   \end{equation}

\begin{equation}
    \Delta F_{ind}^{AR}(t) = a_2 + b_2 \cdot \Delta
F_{10.7}^{AR}(t) \,.
   \end{equation}

The coefficients  $a_1$ and $b_1$ vary from  $a_2$ and $b_2$ in
different power for our different activity indices. For SSN this
difference is small, but for Counts of flares index the difference
between $a_1$, $b_1$ and $a_2$, $b_2$ is more significant than for
SSN.

During the rise and decline cycle's phases the dependence $
F_{ind}(t)$ versus $F_{10.7}(t)$ is approximately linear and
relative addition flux from active regions $\Delta F_{ind}^{AR}(t)$
is neglect with respect to $ F_{ind}^{background}(t)$. So additional
flux from active regions cannot  destroy a balance in the
correlation close to linear  between $ F_{ind}(t)$ and $F_{10.7}(t)$
and respective values of $K_{corr}(t)$ reach their maximum values
during all over the cycle.

During the minimum of activity cycle both values
$F_{ind}^{background}(t)$ and $\Delta F_{ind}^{AR}(t)$ are small,
but additional flux from active regions is not neglect in relation
to background flux that has the minimum values during all over the
cycle.

During the maximum of activity cycle $\Delta F_{ind}^{AR}(t)$ often
exceeds $F_{ind}^{background}(t)$ so disbalance in linear regression
between activity indices increases and values of $K_{corr}(t)$ also
reach their minimum values during all over the cycle too.

\vskip12pt
\centerline {\bf6. Conclusions}
\vskip12pt

For a long time the scientists were interested in the simulation of
processes in the earth's ionosphere and upper atmosphere. It's known
that the solar radiance at $30.4$ nm is very significant for
determination of the Earth high thermosphere levels heating.
(Lukyanova \& Mursula 2011) showed that the for solar $30.4$ nm
radiance fluxes forecasts (very important for Earth thermosphere's
heating predictions) there were more prefer to use Mg II $280$ nm
observed data.

Although $F_{10.7}$ does not actually interact with the Earth
atmosphere $F_{10.7}$ is a useful proxy for the combination of
chromospheric, transition region and coronal solar EUV emissions
modulated by bright solar active regions whose energies at the Earth
are deposited in the thermosphere (Tobiska {\it et al.} 2008).
$F_{10.7}$ dependence on few processes, combined with it localized
formation in the cool corona, i.e. region that is closely coupled
with magnetic structures responsible for creating the XUV-EUV
irradiances, make this a good generalized solar proxy for
thermospheric heating.

 (Tobiska {\it et al.} 2008) presented the improved thermospheric density model, where four solar and
 two geomagnetic indices were used. Solar indices are $F_{10.7}$, 26-34 nm EUV emission,
 Mg II core-to-wing ratio, X-rays in the 0.1-0.8 nm. The geomagnetic
 indices are ap index (amplitude of planetary geomagnetic activity - which
 is derived from geomagnetic field measurements made at several locations
 around the world) and Dst index (Disturbance Storm Time - as indicator
 of the storm-time ring current in the inner magnetosphere). It was
 proved the efficiency of  of the simultaneous use of multiple indexes of solar and geomagnetic activity.

In this paper we found out the cyclic behavior of a calculated
during three-year interval values of correlation coefficients
$K_{corr}(t)$ of linear regression for TSI, Flare Index, Mg II
$280$ nm and Counts of flares versus $F_{10.7}$ and SSN during
solar activity cycles 21,22 and 23 (see Figure 4). We showed that
$K_{corr}(t)$ have the maximum values at the rise and decline
phases - the linear connection between indices is more strong in
these cases. It means that the forecasts of solar indices, based
on $F_{10.7}$ observations will be more successful during the rise
and decline cycle's phases.  We showed that the linear correlation
of activity indices $F_{530}$ , Mg II index, Flare Index and TSI
versus $F_{10.7}$ is stronger  than the linear correlation of
these indices versus SSN but the linear correlation  of Counts of
flares has no difference versus $F_{10.7}$ or versus SSN. This may
be due, in particular, that all indices have the  "Quiet Sun
Level" different from zero except SSN, which has the minimum value
equal to zero.

We also determined that a calculated during three-year interval
values of correlation coefficients $K_{corr}(t)$ show (for the
linear regressions assumption) that for the solar indices versus
$F_{10.7}$ and SSN the minimum values were achieved two times during
the 11-year cycle.

Our study of linear regression between solar indices and $F_{10.7}$
confirms the fact that at minimum and at maximum cycle's phases the
nonlinear state of interconnection between solar activity indices
(characterized the energy release from different layers of solar
atmosphere) increases.

\bigskip
{\bf Acknowledgements} The authors thank the RFBR grant 12-02-00884
 for support of the work.

\bigskip
{\bf References}
\bigskip

Alekseev, I.Yu. \& Gershberg, R.E. 1996, On spotting of red dwarf
stars: direct and inverse problem of the construction of zonal
model, {\it Astronomy Report}, {\bf 73}, 589.

Baliunas, S.L., \& Donahue, R.A., \& Soon, W.H. et al. 1995,
Chromospheric variations in main-sequence stars, {\it Astrophysical
Journal}, {\bf  438}, 269.

Bowman, B.R, \&  Tobiska, W.K., \& Marcos, F.A. et al., 2008,  A New
Empirical Thermospheric Density Model JB2008 Using Solar and
Geomagnetic Indices, {\it AIAA/AAS Astrodynamics Specialist
Conference, AIAA 2008-6438}.

Bruevich, E.A., \& Nusinov A.A, 1984,  Spectrum of short-wave
emission for aeronomical calculations for different levels of solar
activity, {\it Geomagnetizm i Aeronomia}, {\bf24}, 581.

Bruevich E.A. 1995, $H_{Alpha}$ Line Profile in a Gas-Dynamical
Model of Solar Flares, {\it Astronomy Reports},
 {\bf{39}}, N1, 78.

Bruevich P.V. \& Bruevich E.A. 2004, The Possibility of Simulation
of the Chromospheric Variations for Main-Sequence Stars,
{\it{Astronomical and Astrophysical Transactions}}, {\bf{23}}, Issue
2, p. 165-172. DOI: 10.1080/10556790410001666319

Bruevich, E.A., \& Alekseev I.Yu. 2007, Spotting in stars with a low
level of activity, close to solar activity, {\it  Astrophysics},
{\bf 50}, No 2, 187.

Bruevich, E.A., \& Kononovich E.V. 2011, Solar and Solar-type Stars
Atmosphere's Activity at 11-year and Quasi-biennial Timecales, {\it
Moscow University Physics Bulletin}, {\bf  N1}, 66, 72.

Bruevich, E.A. \& Ivanov-Kholodnyj G.S.. 2011,  On Cyclic Activity
of The Sun and Solar-Type Stars, eprint ArXiv:1108.5432

Bruevich, E.A., \& Yakunina, G.V., 2011,  Solar Activity Indices in
the Cycles 21 - 23, eprint ArXiv:1102.5502v1

Chapman, R.D., \& Neupert, W.M., 1974, Slowly varying component of
extreme ultraviolet solar radiation and its relation to solar radio
radiation,  {\it  J. Geophys. Res.}, {\bf 79}, 4138.

Donnelly, R.F., \& Heath, D.F., \& Lean, J. L. \& Rottman, G.J.,
1983,  Differences in the temporal variations of solar UV flux,
10.7-cm solar radio flux, sunspot number, and Ca-K plage data caused
by solar rotation and active region evolution, {\it J. Geophys.
Res.}, {\bf88}, 9883.

Fligge, M., \& Solanki, S.K., \& Unruh, Y.C., \& Frohlich, C., \&
Wehrli, C. 1998,  A model of solar total and spectral irradiance
variations. {\it  Astronomy \& Astrophys.} {\bf 335}, 709.

Floyd, L., \& Newmark, J., \& Cook, J., \& Herring, L., \& McMullin,
D. 2005, Solar EUV and UV spectral irradiances and solar indices,
{\it Journal of Atmospheric and Solar-Terrestrial Physics}, {\bf
67}, 3.

Gaizauscas, V. \& Tapping, K.F., 1988,  Compact sites at 2.8 cm
wavelength  of microwave emission  inside solar active regions. {\it
Astrophys. J.}, {\bf 325}, 912.

Ishkov, V.N. 2009, {\it 1st Workshop on the activity cycles on the
Sun and stars, Moscow, 18-10 December, Edited by EAAO,
St-Petersburg}, p. 57.

Janardhan, P. \&  Susanta, K.B. \& Gosain, S., 2010 {\it Solar Polar
Fields During Cycles 21 - 23: Correlation with Meridional Flows,
Solar Physics}, {\bf 267}, 267.

Kleczek, J. 1952, Catalogue de l'activite' des e'ruptions
chromosphe'riques. {\it  Publ. Inst. Centr. Astron.}, {\bf 22}.

Krivova, N.A., \& Solanki, S.K., \& Fligge, M., \& Unruh, Y. C.
2003,  Reconstruction of solar total and spectral irradiance
variations in cycle 23: is solar surface magnetism the cause?, {\it
Astron. Astrophys.} {\bf 339}, L1.

Krivova, N. A., \& Solanki, S. K. 2008,  Models of solar irradiance
variations: current status, {\it Journal of Astrophysics and
Astronomy}, {\bf 29}, 151.

Kruger, A. 1979, {\it Introduction to Solar Radio Astronomy and
Radio physics, D. Reidel Publ. Co., Dordrecht, Holland}.

Lean, J. L., 1990, A comparison of models of the Sun's extreme
ultraviolet irradiance variations, {\it  J. Geophys. Res.}, {\bf95},
11933.

Livingston, W., \& Penn, M. J.,  \& Svalgaard  L., 2012, Decreasing
Sunspot Magnetic Fields Explain Unique 10.7 cm Radio Flux, {\it
Astrophys. J.}, {\bf 757}, N1, L8.

Lukyanova, R., \& Mursula, K. 2011,  Changed relation between
sunspot numbers, solar UV/EUV radiation and TSI during the declining
phase of solar cycle 23, {\it
 Journal of Atmospheric and Solar-Terrestrial Physics}
{\bf 73}, 235.

Nagovitsyn, Y.A., \& Pevtsov, A.A., \& Livingston W.C. 2012,  On a
possible explanation of the long-term decrease in sunspot field
strength, {\it Astrophysical Journal Letters}, {\bf 758}, L20.

National Geophysical Data Center. Solar-Geophysical Data Reports. 54
Years of Space Weather Data. 2009, {\it
http://www.ngdc.noaa.gov/stp/solar/sgd.html}.

National Geophysical Data Center. Solar Data Service. Sun, solar
activity and upper atmosphere data. 2013, {\it
http://www.ngdc.noaa.gov/stp/solar/solardataservices.html}.

Nicolet, M., \& Bossy, L., 1985,  Solar Radio Fluxes as indices of
solar activity, {\it Planetary Space Sci.}, {\bf33}, 507.

Penn, M.J., \& Livingston, W.C. 2006,  Temporal Changes in Sunspot
Umbral Magnetic Fields and Temperatures, {\it Astrophysical Journal
Letters}, {\bf649}, L45.

Pevtsov A. A., \& Nagovitsyn, Y. A., \& Tlatov, A. G., \& Rybak, A.
L. 2011, Long-term Trends in Sunspot Magnetic Fields, {\it
 Astrophysical Journal Letters}, {\bf742}, L36.

Rozgacheva I.K. \& Bruevich E.A. 2002,  Model of Laminar Convection
in Solar Type Stars, {\it Astronomical and Astrophysical
Transactions}, {\bf{21}}, Issue 1, p. 27. DOI:
10.1080/10556790215583

Skupin, J., \& Noyel, S.,\& Wuttke, M.W., \& Gottwald, M., \&
Bovensmann, H., \& Weber, M., \& Burrows, J. P. 2005,  SCIAMACHY
solar irradiance observation in the spectral range from 240 to 2380
nm, {\it Advance Space Res.}, {\bf 35}, 370.

Svalgaard, L., \& Lockwood M., \& Beer J. 2011, Long-term
reconstruction of Solar and Solar Wind Parameters, {\it
http://www.leif.org/research/Svalgaard\_ISSI\_Proposal\_Base.pdf}.

Svalgaard, L. \& Cliver E.W. 2010, Heliospheric magnetic field
1835-2009, {\it  J. Geophys. Res.}, {\bf115}, A09111.

Tapping, K.F., \& DeTracey, B., 1990,  The origin of the 10.7 cm
flux, {\it Solar Physics}, {\bf127}, 321.

Tobiska, W.K., \&  Bouwer S.D., \& Bowman, B.R., 2008, The
development of new solar indices for use in thermospheric density
modeling, {\it J. Atmospheric \& Solar-Terrestrial Phys.},{\bf 70},
803.

Viereck, R., \& Puga, L., \& McMullin, D., \& Judge, D., \& Weber,
M. \& Tobiska, K. 2001, The MgII index: a proxy for solar EUV, {\it
 Journal of Geophysical Research} , {\bf 73}, No 7, 1343.

Viereck, R.A., Floyd, L.E., Crane, P.C., Woods, T.N., Knapp, B.G.,
Rottman, G., Weber, M., Puga, L.C. and Deland, M.T. 2004,  A
composite MgII index spanning from 1978 to 2003. {\it Space
Weather}, {\bf 2}, No. 10, doi:10.1029/2004SW000084.

Vitinsky, Yu.,\& Kopezky, M., \& Kuklin G., 1986. {\it The sunspot
solar activity statistik}, Moscow,  Nauka.

\end{document}